
\magnification=1200
\baselineskip=12pt
\rightline{UR-1316\ \ \ \ \ \ \ }
\rightline{ER-40685-766}
\rightline{IC/93/216\ \ \ \ }
\baselineskip=20pt
\centerline{\bf ANNIHILATION DIAGRAMS IN TWO-BODY NONLEPTONIC}
\centerline{\bf DECAYS OF CHARMED MESONS}

\vskip 1cm

\centerline{P. Bedaque$^*$, A. Das$^*$}
\centerline{International Centre for Theoretical Physics}
\centerline{Trieste, Italy}
\centerline{and}
\centerline{V. S. Mathur}
\centerline{Department of Physics and Astronomy$^*$}
\centerline{University of Rochester}
\centerline{Rochester, NY 14627}

\vskip 1.5cm

\noindent{\bf Abstract}

In the pole-dominance model for the two-body nonleptonic decays of charmed
mesons $D \rightarrow PV$ and $D \rightarrow VV$, it is shown that the
contributions of the intermediate pseudoscalar
and the axial-vector meson poles cancel each other in the annihilation
diagrams in the chiral limit.  In the same limit, the annihilation diagrams
for the $D \rightarrow PP$ decays vanish independently.

\vskip 3cm

\noindent $^*$Permanent address

\vfill\eject

In a recent paper$^1$, we studied the two-body nonleptonic decays of charmed
mesons using a simple pole-dominance model involving the vector, pseudoscalar
and axial-vector mesons poles.  A salient feature
 of this calculation in the decay to a final
state of a pseudoscalar and a vector meson (PV) was the destructive
interference between the pseudoscalar and the axial-vector meson pole
contributions to the annihilation diagrams.  In particular, in decays like
$D^0 \rightarrow \phi \overline K^0$ and
$D^+_S \rightarrow \rho^+ \pi^0$, which proceed only through annihilation
diagrams, one finds large cancellations in the contributions of the
pseudoscalar and axial-vector meson poles.  This paper is devoted to a study
of this interference effect.

For nonleptonic decay of charm, the effective weak Hamiltonian may be written
as$^{2,3}$
$$H_W = {G_F \over \sqrt{2}} \left[ a_1 (\overline u d^\prime)_\mu
(\overline s^\prime c)_\mu +
a_2 (\overline s^\prime d^\prime )_\mu (\overline uc)_\mu
\right] \eqno(1)$$
where $(\overline q^\alpha q_\beta)_\mu$ are color-singlet V-A currents
$$\left( \overline q^\alpha q_\beta \right)_\mu = i \overline q^\alpha
\gamma_\mu \left( 1 + \gamma_5 \right) q_\beta =
\left( V_\mu \right)^\alpha_\beta + \left( A_\mu \right)^\alpha_\beta
\qquad \left( \alpha , \beta = 1,2 \dots 4 \right) \eqno(2)$$
and $a_1,\  a_2$ are real coefficients which will be treated as
phenomenological parameters.  The primed quark fields are related to the
unprimed ones by the usual Cabibbo-Kobayaski-Maskawa (CKM) mixing matrix.

In the pole-dominance model, we take the currents in $H_W$ to be the
hadronic currents given by the field-current identities
$(\alpha, \beta =1,2 \dots 4)$.
$$\eqalign{(V_\mu)^\alpha_\beta &= \sqrt{2}\ g_V (\phi_\mu)^\alpha_\beta\cr
(A_\mu)^\alpha_\beta &= \sqrt{2}\ f_P \partial_\mu
 P^\alpha_\beta + \sqrt{2}\ g_A (a_\mu)^\alpha_\beta\cr}\eqno(3)$$
where $(\phi_\mu)^\alpha_\beta,\ P^\alpha_\beta$ and
$(a_\mu)^\alpha_\beta$ are the field operators of the vector,
the pseudoscalar and the axial-vector mesons, respectively, and $g_V,\
 f_P$ and
$g_A$ are the corresponding decay constants.  The nonleptonic weak interaction
can then be represented by a two-meson vertex
obtained on substituting (3) into (1).

The Feynman diagrams for the two-body decays $D \rightarrow PP,\  PV$ and
 $VV$ can be readily
drawn$^1$ in terms of the vector,
 pseudoscalar and axial-vector meson poles.  The
pseudoscalar and axial-vector meson poles can contribute only to $D
\rightarrow  PV$ and
$D \rightarrow VV$, but not to $D \rightarrow PP$.
 In Figs. 1 and 2, we display these contributions to
the annihilation-type Feynman diagrams for $D \rightarrow PV$ and
$D \rightarrow VV$, respectively.  In
these figures, the dark dot represents the weak vertex and the open circle
the strong vertex.  Also the dotted, solid and wavy lines represent the
pseudoscalar, vector and axial-vector mesons, respectively.

The strong vertices appearing in Figs. 1 and 2 are of the type $VPP,\  VPA$,
 $VVP$ and
$VVA$.  As in ref. 1, we use an extended spin-SU(4) symmetry to relate the
$VVP$ couplings to the $VPP$ couplings.  In fact we use a generalization of the
Sakita-Wali interaction Hamiltonian$^4$, which relates the $VPP,\  VVP$
 and $VVV$
couplings
$$\eqalign{H_{\rm str} = ig\ {\rm Tr}\ \bigg( \phi_\mu &P{\buildrel
\longleftrightarrow \over{\partial_\mu}} P -
{2 \over M}\ \varepsilon_{\mu \nu \lambda \rho}P
\partial_\mu \phi_\nu \partial_\lambda
\phi_\rho\cr
&+ {2 \over 3}\ F_{\mu \nu} \phi_\mu \phi_\nu -
{2 \over 9 M^2}\ F_{\mu \nu} F_{\nu \lambda} F_{\lambda \mu} \bigg)
\cr}\eqno(4)$$
Here the trace is over the SU(4) multiplets, $g$ is a coupling constant and
M represents a mass scale.  We identify $M$ with the mass of the decaying
particle and take $g$ to be the $\rho \pi \pi$
 coupling determined from the decay
$\rho \rightarrow 2 \pi$.  In our previous work, we described the $VPA$ vertex
phenomenologically, and neglected the $VVA$ interaction.  In the present
 work, we choose to describe the $VPA$ and $VVA$ vertices also in terms of the
extended spin-SU(4) symmetry.  Following$^4$ the arguments that led to (4),
we obtain the interaction Hamiltonian in this case to be
$$\eqalign{H^\prime_{\rm str} = i g^\prime \ {\rm Tr}\ \bigg(
&M \phi_\mu [P,a_\mu]_- + {1 \over 4M}\ F_{\mu \nu} [P,
F^a_{\mu \nu} ]_- \cr
&- {1 \over 2}\ \varepsilon_{\mu \nu \lambda \rho} \partial_\lambda \phi_\rho
[\phi_\mu , a_\nu ]_+ \bigg)\cr}\eqno(5)$$
where $[X,Y]_\pm$ represent the anticommutator and the commutator,
respectively, $g^\prime$ is a different coupling constant and
$$F_{\mu \nu} = \partial_\mu \phi_\nu - \partial_\nu \phi_\mu \quad , \quad
F^a_{\mu \nu} = \partial_\mu a_\nu - \partial_\nu a_\mu \eqno(6)$$
are the field tensors.

Now, the coupling constants $g$ and $g^\prime$
 can be related through chiral symmetry.
The simplest way to see this is to consider the matrix element of an
axial-vector current between an ordinary vector and
 an ordinary pseudoscalar meson
$<P|A_\mu|V>$.  From Lorentz invariance, this can be written in the form
$$\eqalign{<P(k)|A_\mu (0)|V(p)>\ =\  &{i \varepsilon^V_\nu (p) \over
(4 p_0 k_0 V^2)^{1/2}}\cr
&\big[ K_1 (q^2)
\delta_{\mu \nu} + K_2 (q^2) k_\nu (p+k)_\mu
+ K_3 (q^2) k_\nu (p-k)_\mu \big]\cr} \eqno(7)$$
where $K_{1,2,3}(q^2)$ are the form-factors and $q=p-k$.  Chiral SU(3)
symmetry, with massless pseudoscalar mesons, then implies
$$K_1 (q^2) + m^2_V K_2 (q^2) - q^2 K_3 (q^2) = 0 \eqno(8)$$
We may now use the pole-dominance model to determine the form-factors in
terms of the pseudoscalar and axial-vector meson poles.  The corresponding
Feynman diagrams are shown in Fig. 3.  The dash-and-dot line represents the
axial-vector current and the vertices denoted by crosses can be read
 off from the field-current identity in Eq. (3).  We find
$$\eqalign{K_1 (q^2) &= -2M {g_A g^\prime \over q^2 + m^2_A}\
\big[ 1 + {1 \over 4M^2}\ \big( q^2 - m^2_V \big) \big]\cr
\noalign{\vskip 4pt}%
K_2 (q^2) &= - {1 \over 2M}\ {g_A g^\prime \over q^2 + m^2_A}\cr
\noalign{\vskip 4pt}%
K_3 (q^2) &= {4 f_P g \over q^2} + {2M \over m^2_A}\
{g_A g^\prime \over q^2 + m^2_A} \ \bigg(
1 - {m^2_A \over 4 M^2}\bigg)\cr}\eqno(9)$$
Substituting (9) in (8), we readily obtain
$$2f_P g + {g_A \over  m^2_A}\  M g^\prime = 0 \eqno(10)$$

Returning to the decays $D \rightarrow PV$ and $D \rightarrow VV$,
we define the decay amplitudes as follows
$$M \left( D(q)
 \rightarrow V_1 \left( q_1 , \lambda_1 \right)P_2 \left( q_2 \right)
\right) = {-i (2 \pi)^4 \delta^{(4)} (q -q_1 - q_2 ) \over
\sqrt{2q_0 V 2q_{10} V 2q_{20} V}} q_2 \cdot \varepsilon^{(\lambda_1)}
(q_1) B \eqno(11)$$
$$\eqalign{M \big( D(q)
 &\rightarrow V_1 ( q_1 , \lambda_1)V_2 ( q_2,
\lambda_2)
\big) = {-i (2 \pi)^4 \delta^{(4)} (q -q_1 - q_2 ) \over
\sqrt{2q_0 V 2q_{10} V 2q_{20} V}}\cr
\noalign{\vskip 4pt}%
&\big[ i C \delta_{\alpha \beta} +
i D \varepsilon_{\mu \alpha \nu \beta} q_{1 \mu} q_{2 \nu}
+ i E q_{1 \beta} q_{2 \alpha} \big]
\varepsilon^{(\lambda_1)}_\alpha (q_1)
\varepsilon_\beta^{(\lambda_2)} (q_2)\cr}\eqno(12)$$
The contributions of the pseudoscalar and axial-vector meson poles to the
annihilation diagrams in Fig. 1 for the decay
$D \rightarrow PV$ are easily calculated to
be
$$\eqalign{B_{\rm ann} (P) &= - 4 \ \alpha\  a\  f_D f_P\  {m^2_D \over
m^2_P - m^2_D}\ g\cr
\noalign{\vskip 4pt}%
B_{\rm ann} (A) &= 2 \ \alpha\ a\ f_D
 {g_A \over m^2_A} M g^\prime\cr}\eqno(13)$$
where $a$ stands for the coefficient $a_1$ or
$a_2$ in the weak Hamiltonian (1),
depending on which term in (1) contributes, and
$$\alpha = {G_F \over \sqrt{2}}\ V^*_{cs} V_{ud} \eqno(14)$$
In (14), the $V$'s represent the matrix elements of the $CKM$ matrix.  Here we
are considering the Cabibbo allowed decays, but the considerations can be
readily extended to the Cabibbo suppressed decays also.  Note that in Fig.
1, $P$ is an ordinary pseudoscalar meson, so that in the chiral limit with
$m_P \rightarrow 0$, we get from (13), on using the result (10)
$$B_{\rm ann} (P) + B_{\rm ann} (A) = 0 \eqno(15)$$
Thus, in the chiral limit, the pseudoscalar and axial-vector meson pole
terms cancel each other in the annihilation diagrams of the decay
$D \rightarrow PV$.

For the decay $D \rightarrow VV$, we similarly find the contributions of
the pseudoscalar and axial-vector meson poles to the annihilation diagrams
in Fig. 2 to be
$$\eqalign{C_{\rm ann} (P) &= E_{\rm ann} (P) = 0\cr
D_{\rm ann} (P) &= 4\  \alpha \ a\ f_D f_P \
{m^2_D \over m^2_P - m^2_D}\ {g \over M}\cr
C_{\rm ann} (A) &= E_{\rm ann} (A) = 0\cr
D_{\rm ann} (A) &= - 2 \ \alpha \ a\ f_D {g_A \over m^2_A}
 g^\prime\cr}\eqno(16)$$
Once again in the chiral limit, we find on using (10)
$$D_{\rm ann}(P) + D_{\rm ann} (A) = 0 \eqno(17)$$
Thus, in the chiral limit, the pseudoscalar and axial-vector meson pole
contributions to the annihilation diagrams in $D \rightarrow VV$
also cancel each other.

It should also be noted that in the decay $D \rightarrow P_1 P_2$,
where only the vector meson pole contributions, the annihilation diagram
also vanishes in the chiral limit.  This is easily seem from the result$^5$
$$A_{\rm ann} \propto {m^2_2 - m^2_1 \over m^2_V} \eqno(18)$$
for the vector meson contribution to the annihilation amplitude in
$D \rightarrow P_1 P_2$, where $m_1,\ m_2$ are the masses of the
pseudoscalar mesons in the final state.

It should be remarked that the vanishing of the annihilation amplitudes in
the chiral limit is not surprising.  Indeed, it is the analogue of the
helicity suppression of the annihilation amplitude in the quark model.  In
fact, for massless quarks (in the chiral limit),
 it is well-known that the quark model
annihilation amplitude vanishes.  Also our result is consistent with the
result of Bauer et al.$^3$ for the suppression of annihilation amplitude in
the factorization model.

With $g^\prime$ determined in the chiral limit from Eq. (10),
 we could calculate all two-body decays of the $D$ mesons in the
pole-dominace model$^1$ just in terms of the parameters $a_1$
 and $a_2$ appearing in the weak Hamiltonian (1).  However, in view of the
sensitivity of the destructive interference effect discussed above
 to small variations in $g^\prime$, this is not the best approach.  It
should be noted that uncertainties in the determination of $g^\prime$ from
  Eq. (10) arise not only due to the use of chiral symmetry
 but also the choice of the mass scale $M$.  In view of this, the
phenomenological approach to the calculation of the decay
 rates followed in ref. 1  is preferable.

Finally, we can calculate the width for the decay $A_1 \rightarrow \rho
 \pi$
which also depends on $g^\prime$.  Using (10), we find
$g^\prime = -6.5$, which gives
$$\Gamma (A_1 \rightarrow \rho \pi ) = 566 \ {\rm MeV} \eqno(19)$$
Experimentally, this width is not well-determined but our result may be
compared with the value quoted in the particle properties data booklet$^6$
 of $\sim$400 MeV.

\noindent {\bf Acknowledgement}

This work was supported in part by the U.S. Department of Energy Grant No.
DE-FG-02-91ER40685.

\vfill\eject
\noindent {\bf References and Footnotes}
\medskip
\item{1.} P. Bedaque, A. Das and V. S. Mathur, University of Rochester
preprint no. UR-1314 (1993).

\item{2.} K. Jagannathan and V.S. Mathur, Nucl. Phys. {\bf B171}, 78 (1980).

\item{3.} M. Bauer, B. Stech and M. Wirbel, Z. Phys. C - Particles and Fields
{\bf 34}, 103 (1987).

\item{4.} B. Sakita and K. C. Wali, Phys. Rev. Lett. {\bf 14}, 404 (1965) and
Phys. Rev. {\bf 139}, B1355 (1965); A. Salaam, R. Delbourgo and J.
Strathdee, Proc. Roy. Soc. (London) {\bf 284}, 146 (1965).

\item{5.} A. Das and V. S. Mathur, Mod. Phys. Lett. A {\bf 8},
 2079 (1993). Note that
(18) also vanishes in the SU(3) symmetric limit.
\smallskip

\item{6.} Review of Particle Properties, Phys. Rev.
{\bf D45}, Part 2 (June 1992).
\vfill\eject

\noindent {\bf Figure Captions}
\medskip

\item{\bf Fig. 1} Feynman diagrams for the pseudoscalar and axial-vector
meson pole contributions to the annihilation amplitude in the decay
$D \rightarrow PV$.

\item{\bf Fig. 2} Feynman diagrams for the pseudoscalar and
axial-vector meson meson pole contributions to the annihilation
amplitude in the decay $D \rightarrow VV$.

\item{\bf Fig. 3} Feynman diagrams for the pseudoscalar and
 axial-vector meson pole contributions to the form-factors in Eq. (7).

\end